\documentclass{article}
\usepackage{graphicx}
\begin{document}
\title{Higher-order vortex solitons, multipoles, and supervortices on a
square optical lattice}
\author{Hidetsugu Sakaguchi$^1$ and Boris A. Malomed$^2$\\
$^1$ Department of Applied Science for Electronics and Materials,\\
Interdisciplinary Graduate School of Engineering Sciences,\\
Kyushu University, Kasuga, Fukuoka 816-8580, Japan\\
$^2$ Department of Interdisciplinary Studies,\\
School of Electrical Engineering, Faculty of Engineering,\\
Tel Aviv University, Tel Aviv 69978, Israel}
\maketitle
\begin{abstract}
We predict new generic types of vorticity-carrying soliton
complexes in a class of physical systems including an attractive
Bose-Einstein condensate in a square optical lattice (OL) and
photonic lattices in photorefractive media. The patterns include
ring-shaped higher-order vortex solitons and supervortices.
Stability diagrams for these patterns, based on direct
simulations, are presented. The vortex ring solitons are stable if
the phase difference $\Delta \phi $ between adjacent solitons in
the ring is larger than $\pi /2$, while the supervortices are
stable in the opposite case, $\Delta \phi <\pi /2$. A qualitative
explanation to the stability is given.
\end{abstract}

\bigskip
Recent experimental observations of two-dimensional (2D) solitons
\cite{Moti} and localized vortices \cite{vortex} in photonic
lattices in self-focusing photorefractive crystals (PRCs), and
prediction of similar structures in the 2D Gross-Pitaevskii
equation (GPE)\ for a self-attractive Bose-Einstein condensate
(BEC) loaded into the square optical lattice (OL)
\cite{BBB1,JiankeZiad}, have drawn interest to solitons of this
type, including vorticity-carrying ones. Vortex solitons of the
gap type have also been predicted in the repulsive GPE with the
square OL \cite{Ostr,we}, including higher-order ones, with the
vorticity $S=2$ \cite{we}. However, the results which were
reported thus far for the self-focusing media, both saturable and
cubic, pertain solely to the vortices with $S=1$ (including
strongly deformed ones \cite{Canberra}). In this work, we present
stable vortex solitons with higher values of $S$ and more complex
vortex patterns, which suggests a possibility to create them in
the experiment. We report the results for both BEC and PRC models,
which attests to their generic character. It is relevant to
mention that, in discrete models, which may be considered as a
limiting case corresponding to a very strong OL, with attraction
and repulsion alike, both 2D and 3D vortex solitons with $S=2$ are
unstable, but the ones with $S=3$ have their stability region
\cite{Panos}.

In a moderately strong OL, we find stable vortex solitons with
higher vorticities $S$ and their counterparts in the form of
quadrupole solitons (in the discrete model, the latter ones may be
stable too \cite{Panos}). The vortices with large $S$ feature a
ring shape, resembling vorticity-carrying ``soliton necklaces",
that were recently studied in diverse spatially uniform
\cite{necklace} models, but, on the contrary to those models, the
ring-shaped vortices on the square lattice are true stationary
states, rather than slowly disintegrating quasi-patterns in the
free space (recently, zero-vorticity necklaces and other soliton
complexes were predicted and observed in Kerr and PRC media with a
photonic lattice \cite{Zhigang}). We also construct another
generic type of stable structures that we call ``supervortices"
\cite{superfluxon}, i.e., ring-shaped (or densely packed) arrays
of individual compact vortices with $s=1$, arranged so that the
ring itself carries global vorticity ($s$ pertains to the
intrinsic vorticity of an individual vortex in the array, to
distinguish it from $S$ referring to the global vorticity
imprinted into the array). Although the ring-shaped supervortices
may formally resemble dark-soliton vortex necklaces in external
traps \cite{dark-vortex-necklace}, they have never been considered
before, as the individual solitary-wave vortex in unstable without
the OL.

In the normalized form, the two-dimensional (2D) GPE with the negative
atomic scattering length and square-OL potential is \cite{BBB1}
\begin{equation}
i\psi _{t}+(1/2)\nabla ^{2}\psi +|\psi |^{2}\psi +\varepsilon \left[ \cos
(2x)+\cos (2y)\right] \psi =0,  \label{GP}
\end{equation}where $\psi (x,y,t)$ is the BEC wave function, the OL period is scaled to be
$\pi $, and $\varepsilon $ is the OL strength, in units of the
respective recoil energy. The vertical coordinate is excluded,
assuming tight confinement in the third direction, which is
readily provided by light sheets \cite{LightSheet}. As is known
(see, e.g., Ref. \cite{PLA}), the underlying three-dimensional GPE
with a negative scattering length of atomic collisions, $a<0$, can
be reduced to its 2D counterpart (\ref{GP}), provided that the
transverse quantum pressure is much stronger than the
self-attraction (which precludes the collapse). In other words,
the energy of the atom with the mass $m$ in the ground state of
the vertical trap with a size $a_{z}$, $E_{0}^{(z)}\simeq \hbar
^{2}/\left( 2ma_{z}^{2}\right) $, must be much larger than the
contribution of the attraction between atoms to the atomic
chemical potential, $\Delta \mu \simeq -2\pi \hbar ^{2}\left(
a/m\right) n,$where $n$ is the three-dimensional density of atoms.
Thus, the 2D GPE equation for the self-attractive condensate is
relevant under the condition $n\ll \left( 4\pi a_{z}^{2}a\right)
^{-1}$, which can be readily met in the experiment (in particular,
$a$ can be very small in $^{7}$Li \cite{Li}). In fact, this
condition can be relaxed due to the stabilizing effect of the OL
potential in Eq. (\ref{GP}).

Equation (\ref{GP}) does not include the external trapping
(parabolic) potential, as we aim to look for patterns whose
stability does not depend on the presence of the trap. An equation
governing the spatial evolution of the amplitude $\psi (x,y,z)$ of
the probe beam in the PRC with a square photonic lattice of
strength $I_{0}$ differs from Eq. (\ref{GP}) by the presence of
saturation \cite{Moti,Yang},\begin{equation} i\psi
_{z}+(1/2)\nabla ^{2}\psi +\frac{\psi }{1+I_{0}[\cos (Kx)+\cos
(Ky)]^{2}+|\psi |^{2}}  \label{PL}
\end{equation}(strictly speaking, PRCs feature anisotropy, but in crystals used in work
\cite{Moti} the anisotropy is very weak and may be neglected).
Stationary solutions with chemical potential $\mu $ are looked for
as $\psi (x,y,t)=e^{-i\mu t}\Psi (x,y)$, where the function $\Psi
$ is, generally, complex (in the PRC model, $-\mu $ is the
propagation constant). In this notation, the solutions depend on
two parameters: the lattice strength, $\varepsilon $ or $I_{0}$,
and the norm (number of atoms/total power of the light signal),
$N=\int \int |\Psi (x,y)|^{2}dxdy$.

Equations (\ref{GP}) and (\ref{PL}) represent a class of models that extend
to other physical media, such as nonlinear photonic crystals and
photonic-crystal fibers \cite{PhotCryst}. The results reported below are
valid for all these models, i.e., they suggest, in particular, the existence
of stable vortex complexes of the new types in the photonic crystals as well
(a vortex with $s=1$ was very recently predicted in photonic-crystal fibers
\cite{PhotCrystFiber}).

We built patterns with vorticity $S$ as sets of $M$ stable
``single-cell" solitons with $s=0$, or compact vortices with
$s=1$, by imprinting onto the set a phase distribution with the
phase shift $\Delta \phi =2\pi S/M$ between adjacent solitons. The
stability of the resulting states was tested in direct
simulations. Initial shapes of the building blocks with $s=0$ and
$s=1$ were taken as numerical solutions of the axially symmetric
GPE with an \textit{ad hoc} potential, $U(r)=-\varepsilon\cos(2r)$
for $0<r<\pi/2$, and $U(r)\equiv\varepsilon$ for $r>\pi/2$, where
$r\equiv\sqrt{x^2+y^2}$, and $\varepsilon$ is the same as in Eq.
(\ref{GP}) And small random perturbations were further added. In
cases of stabilities, it was observed that the shapes and
structures of the vortex and supervortex composed of the building
blocks changed hardly in time. It implies the vortex and
supervortex structures are rather stable, since our initial
conditions are slightly deviated from the exact stationary
solutions. On the contrary, the vorticity-carrying phase pattern
in unstable states is quickly replaced by an evidently chaotic
distribution of phase, and shows no trend to relax into any simple
configuration. The phase instability produces little effect on the
amplitude shape of the pattern. That is, the profile of $|\phi|$ has 
hardly changed in time. 

\begin{figure}[tbh]
\begin{center}
\includegraphics[height=3.5cm]{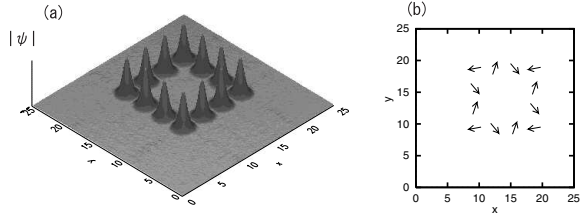}
\end{center}
\caption{(a) A stable vortex soliton with $S=4$ and $N=33.4$;
the field $|\psi (x,y)|$ is plotted. (b) The field of the local
phase vector, $\left( \mathrm{Re}\{\psi \}/|\psi
|,\mathrm{Im}\{\psi \}/|\psi |\right) $, at centers of the
constituent solitons for the vortex complex in (a). In all the
cases shown here, $\protect\varepsilon =3$.} \label{fig1}
\end{figure}

Following this procedure, a stable vortex in Eq. (\ref{GP}) with
 $S=1,2,3$ and $4$ were created
using eight solitons surrounding an empty central cell. We have
found that a vortex with $S=3$ is definitely stable, the one with $S=2$ is marginally stable.
The "marginal stability" features  the very slowly (linearly) growing perturbation in the phase configuration. Further, vortex solitons were also
constructed based on $12$ single-cell solitons surrounding a
cluster of four empty cells. In the latter case, the vortices are
stable for $S=4,5,$ and $6$. A stable vortex with $S=4$ is shown
in Figs. \ref{fig1}(a) and (b). 
 Monitoring the rotation of the
local phase vector, which is defined as one with the components
$\left( \mathrm{Re}\{\psi \}/|\psi |,\mathrm{Im}\{\psi \}/|\psi
|\right) $, along a closed path in the counter-clockwise direction
in Fig. \ref{fig1}(b), it is easy to see that the corresponding
net change of the field's phase is indeed $\Delta \phi =8\pi $,
corresponding to $S=4$. 

\begin{figure}[tbh]
\begin{center}
\includegraphics[height=3.5cm]{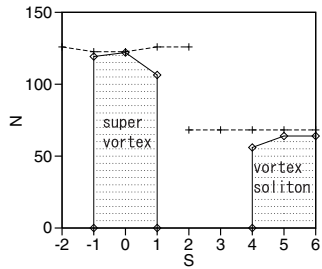}
\end{center}
\caption{The right side of the diagram shows stability limits for
the vortex solitons with higher vortices of the type shown in Fig.
\protect\ref{fig1}(c), for $\protect\varepsilon =3$. Above the
critical norm (crosses), each soliton in the patterns collapses.
The rings are stable in the shaded areas below the borders shown
by rhombuses. In particular, the vortex with $S~=4$ is definitely
stable, while the one with $S~=3$ is marginally unstable. The left
side shows the same limits for densely-packed supervortices of the
type displayed in Fig. \protect\ref{fig3}, with
$\protect\varepsilon =10$. The supervortices are stable in the
shaded area, which actually includes only $S~=\pm 1$ and $S=0$.}
\label{fig2}
\end{figure}

The existence and stability of these solitons depend on $S$, norm
$N$, and $\varepsilon $. Keeping $\varepsilon =3$, and varying
both $N$ and $S$, in the right part of Fig. \ref{fig2} we
summarize stability results for the vortex solitons of Eq.
(\ref{GP}) based on the ring-shaped array of $12$ single-cell
solitons. For the judgement of the stability, we have calculated
the time evolution of the maximum amplitude of $|\psi|$ to check
the collapse, and the time evolution of the root mean square of
$\Delta \phi-2\pi S/12$ for the 12 single-cell solitons to
investigate the phase instability, where $\Delta \phi$ is the
phase difference between adjacent solitons. The numerical
simulation was performed until $t=150$. We have constructed the
stability diagram for the vortex solitons in Fig.~2 from these two
kinds of time evolutions. For all $S$, the pattern collapses (each
individual soliton suffers, as usual, catastrophic
self-compression) if $N$ exceeds a critical value,
$N_{\mathrm{cr}}=68.2$, that virtually does not depend on $S$.
This can be easily understood, taking into regard that, in the
present notation, the critical norm for the onset of the collapse
of an individual soliton in the 2D free-space nonlinear
Schr\"{o}dinger equation is
$N_{\mathrm{cr}}^{(\mathrm{NLS)}}\approx \allowbreak 5.84$
\cite{Berge'}, hence for the set of $12$ pulses it predicts
$N_{\mathrm{cr}}\approx \allowbreak 70.\allowbreak 1$. The small
difference from the actual value, $N_{\mathrm{cr}}=68.2$, is due
to the OL. For $N<N_{\mathrm{cr}}$, all the patterns with $S<3$
are unstable, while the one with $S=3$ is marginal.

The above results for the vortices based on the ring-shaped arrays
of $4$, $8 $ and $12$ single-cell solitons suggest that the
stability is not determined by the vorticity $S$ itself, but
rather by the phase difference $\Delta \phi $ between adjacent
cells. All the vortex complexes built as described above are
stable for $\Delta \phi >\pi /2$ (including the quadrupoles, for
which $\Delta \phi =\pi $), being marginal for $\Delta \phi =\pi
/2$. These inferences can be easily explained. Indeed, it is well
known that sufficiently separated 2D solitons with the phase
difference $\Delta \phi $ interact repulsively if $\Delta \phi
>\pi /2$, and attractively if $\Delta \phi <\pi /2$. On the other
hand, it is also known that bound states of lattice solitons may
be stable only in the case of repulsion \cite{Panos2}. Thus, the
ring arrays trapped on the lattice have a chance to be stable just
in the case of $\Delta \phi >\pi /2$.

It may be relevant to look at several fundamental vortices with $S=1$ studied before from the same perspective.  Baizakov et al. found the fundamental vortex with $S=1$ , which is arranged
as a complex of four major and four minor peaks surrounding a
central cell of the OL (see Fig.3(b) in \cite{BBB1}) and they showed that the vortex is stable by direct numerical analysis \cite{BBB1}. Alexander et al. found numerically a (stable) vortex with $S=1$ constructed of 4 single cell solitons for the PRC model \cite{Canberra}, where the phase difference is $\Delta \phi=\pi/2$. Yang studied the linear stability of the vortex with $S=1$ and found that the vortex is stable in an intermediate range of peak amplitude and it is unstable outside of the parameter range \cite{Yang}. We have found that the vortex with $S=1$ constructed of 4 single cell solitons is marginally stable (very weakly unstable) judging from the time evolution of the phase configuration. However, the amplitude profile $|\psi|$ keeps the initial profile stably and the vortex structure is not broken until $t=150$ for the parameter value of $\varepsilon=3$. In any way, the marginal case of $\Delta \phi=\pi/2$ seems to be delicate and might need further investigation. However, we do not study it more in detail in this paper, because the main topics are higher-order vortex solitons and supervortices.
 
\begin{figure}[tbh]
\begin{center}
\includegraphics[height=3.5cm]{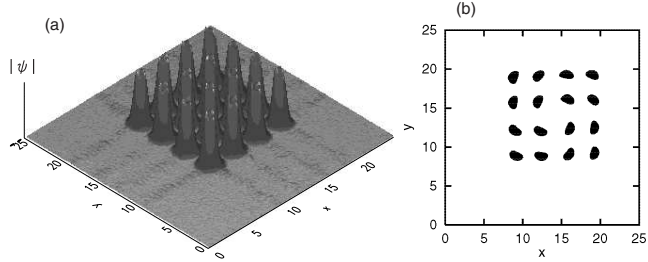}
\end{center}
\caption{A densely-packed stable supervortex with
$\protect\varepsilon =10$, $S=-1$ and $N=74.01$. (a) The plot of
$|\psi (x,y)|$. (b) The area with $\mathrm{Re}\left\{ \psi
(x,y)\right\} >0.2$ is shaded, to highlight the vortex structure
of the pattern, corresponding to $S=-1$.} \label{fig3}
\end{figure}
\begin{figure}[tbh]
\begin{center}
\includegraphics[height=3.5cm]{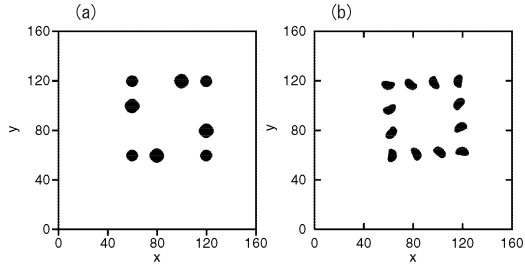}
\end{center}
\caption{Stable vortex patterns in the photorefractive model
(\protect\ref{PL}), with $I_{0}=0.5$ and $K=\protect\pi /10$. (a)
A vortex soliton with $N=438.7$ $\ $and $S=4$. (b) A supervortex
with $N=938.7$ and $S=-1$. The area with $\mathrm{Re}\left\{
\protect\psi (x,y)\right\} >0.1$ is shaded, to highlight the
vortex structures.} \label{fig4}
\end{figure}

Next, we investigate supervortices. To this end, we built a closed
array of $12$ compact vortices with $s=1$ surrounding an empty
cluster of four cells, imprinting the global vorticity onto it the
same way as above (unlike the free space, where the vortex
solitons is always unstable against azimuthal perturbations, in
the lattice it may be completely stable, if taken below the
collapse threshold \cite{BBB1,JiankeZiad}). The resulting
stationary patterns are, to the best of our knowledge, the first
example of ``supervortex" structures on lattices. On the contrary
to the vortex rings considered above, the supervortices
corresponding to $S$ and $-S$ are \emph{not} equivalent.  
In the same case, the supervortices with $S=-1$ and $|S|~=2$ are stable too,
the ones with $|S|~>3$ are unstable, and the supervortices with
$|S|~=3$ are marginal. We have also constructed the simplest
supervortices, composed of four small vortices which cover four
adjacent cells of the square lattice (not shown here). It was
found that such arrays are stable with $S=0$, unstable with
$|S|~=2$ (actually, these are \textit{vortex quadrupoles}), and
marginal with $S=\pm 1$.

The stability of the supervortices composed of $12$ or $4$
individual vortex solitons suggest that all the supervortices with
the phase difference $\Delta \phi $ between adjacent individual
vortices are \emph{unstable} in the case of $\Delta \phi >\pi /2$,
and they may be stable if $\Delta \phi <\pi /2$, exactly opposite
to the vortex rings considered above. This property can be
understood too: the sign of the interaction between compact vortex
solitons differs from that for their zero-spin counterparts by the
factor of $(-1)^{s}$ \cite{me}. Therefore, contrary to the
solitons with $s=0 $ in the ring patterns considered above, the
adjacent vortices with $s=1$ in the supervortex patterns
\emph{repel} each other, provided that $\Delta \phi <\pi /2$,
which gives them a chance to form stable bound states on the
lattice, as per Ref. \cite{Panos2}.

Supervortices may also exist in a densely-packed form, without the inner
hole. As shown in Fig. \ref{fig3}(a), it is possible to build such
structures, starting with a dense set of $16$ compact vortex solitons. The
phase pattern corresponding to the global vorticity $S$ was imprinted, in
this case, by setting $\Delta \phi =\pi S/2$ in the inner layer (formed of
four cells), and $\Delta \phi =\pi S/6$ in the outer one, that includes $12$
cells. Results of the stability investigation for the densely packed
supervortices are included in Fig. \ref{fig2}. In particular, the collapse
threshold (above which each individual vortex collapses independently) is
nearly constant, $N_{\mathrm{cr}}^{(\mathrm{super})}\approx 125$. This value
can be explained similar to how it was done above for the patterns composed
of the $s=0$ solitons, taking into regard the known critical norm for the
collapse of an individual vortex soliton with $s=1$ \cite{Vlasov}. The
densely packed supervortices with $|S\geq 2|$ are unstable.

The PRC model (\ref{PL}) gives rise to vortex solitons and supervortices
very similar to those found in the GPE (\ref{GP}), obeying similar
phase-stability conditions (of course, there is no collapse in the PRC model
with the saturable nonlinearity). Figures 4(a) and (b) display a
higher-order vortex soliton with $S=4$, and a supervortex with $S=-1$. They
are based on $12$ single-cell $s=0$ or $s=1$ solitons which surround four
empty central cells.

To summarize, we have investigated new types of vortex complexes in a class
of models that includes attractive BEC in square optical lattices (OLs) and
photorefractive crystals (PRCs) with the photonic lattice. The same patterns
are expected in nonlinear photonic crystals and photonic-crystal fibers. The
complexes include densely packed and ring-shaped higher-order vortices and
supervortices. They are stable provided that the OL is strong enough.
Stability limits for the total norm and vorticity of the patterns have been
found and explained. The experimental creation of these vortex complexes in
PRCs is quite feasible.

A larger variety of the complex vortex patterns is expected on
triangular and hexagonal lattices, and, especially, in the 3D BEC
model with the corresponding OL, cf. the findings made in the
discrete 3D model \cite{Panos}. A vast ``zoo" of 3D vortex
patterns was very recently reported in Ref. \cite{Spain}, but they
were formed of dark-soliton vortex filaments in the repulsive GPE.

We appreciate discussions with B. B. Baizakov, M. Salerno, and M. Segev. The
work of B.A.M. was supported, in a part, by the Israel Science Foundation
through the grant No. 8006/03.

\end{document}